# Deterministic and probabilistic deep learning models for inverse design of broadband acoustic cloak

Waqas W. Ahmed, Mohamed Farhat, Xiangliang Zhang[*], and Ying Wu[†]

*Division of Computer, Electrical and Mathematical Science and Engineering, King Abdullah University of Science and Technology (KAUST), Thuwal, 23955-6900, Saudi Arabia*

Email: [†]ying.wu@kaust.edu.sa , [*]xiangliang.zhang@kaust.edu.sa

**Abstract:**

Concealing an object from incoming waves (light and/or sound) remained a science fiction for a long time due to the absence of wave-shielding materials in nature. Yet, the invention of artificial materials and new physical principles for optical and sound wave manipulation translated this abstract concept into reality by making an object acoustically '*invisible*'. Here, we present the notion of machine learning driven acoustic cloak and demonstrate an example of such a cloak with a multilayered core-shell configuration. Importantly, we develop deterministic and probabilistic deep learning models based on autoencoder-like neural network structure to retrieve the structural and material properties of the cloaking shell surrounding the object that suppresses scattering of sound in a broad spectral range, as if it was not there. The probabilistic model enhances the generalization ability of design procedure and uncovers the sensitivity of the cloak's parameters on the spectral response for practical implementation. This proposal opens up new avenues to expedite the design of intelligent cloaking devices for tailored spectral response and offers a feasible solution for inverse scattering problems.

**Introduction:**

Cloaking or invisibility, physically related to the cancelation of the natural scattering signature of an object, has attracted enormous attention. In order to render an object invisible to electromagnetic and/or acoustic waves, it is essential to tailor the interaction of waves with the object such that the wave fronts in the surrounding medium remain the same, regardless of the presence of an object. One promising route to achieve invisibility is transformation optics, relying on artificially structured materials[1] that can mold the flow of waves around the concealed object with specific design of its constitutive parameters. However, such transformation-based cloaks[2-4] have the fundamental limitation of narrowband operation, due to the strong dispersion, inherent to the resonance-based meta-atoms, and the undesired material loss and thus challenging due to the difficulty of creating bulky material compositions, with both anisotropy and inhomogeneity[4,5]. In such scenarios, other invisibility schemes based on scattering cancellation technique[6,7], patterned metasurfaces[8,9], and complex modulated potentials[10,11] have also been suggested for different kinds of waves ranging from microwave[4,12,13], acoustic[14-16], elastic[17], and heat waves[18].

In the last decade, acoustic cloaks via scattering cancellation[19,20] has become a topic of interest due to their robust designs, operating spectral range and ease of fabrication. In such schemes, isotropic layers of specific thickness, mass density and bulk modulus can be carefully tailored to cancel the first few scattering orders, which significantly reduce the scattering cross-section of the system, to make the object nearly-undetectable at a particular frequency. Consequently, the scattering cancellation approach generally employs acoustic metamaterials to realize on-demand cloaking devices. However, practical applications often desire more flexibility in the operating frequency band and require designing materials with positive physical properties (density and bulk modulus). Yet, the design of cloaking shells operating over broad frequency ranges with realistic material parameters remains a challenging endeavor. For instance, the broadband cloaking operation requires some additional layers in the core-shell configuration to cancel the higher scattering orders and, as a consequence, the design complexity grows and thus makes it extremely challenging to tune the geometry and material properties with conventional optimization techniques[21]. To mitigate such issues, data-driven approaches based on machine learning have provided a promising platform where artificial neural networks are trained to intelligently learn the intrinsic relation between various structural parameters and their spectral responses, and significantly reduce the overall computational time by predicting the solution immediately after the training phase[22,23].

With the rapid development in machine learning technology, research groups have been able to efficiently solve numerous physical problems including quantum physical problems[24], modulation instability in optical fibers[25], pattern recognition of photonic modes[26], and streamline the inverse design process of protected edge states[27], metasurfaces[28,29], and complex photonic structures for different applications[30-41]. The inverse design process allows for fast and accurate prediction of the design parameters (structure and material properties) with complex architectures such as deep neural networks (DNNs)[29,30], convolutional neural networks (CNNs)[32], recurrent neural network (RNNs)[38], and generative adversarial networks (GANs)[27]. Despite such significant advancement in this area, the reported studies to date mostly emphasize on solving inverse electromagnetic problems in a deterministic fashion, while robust deep learning models for inverse acoustic scattering problems are yet to be developed. Here, we propose deep learning models as a practical tool to design broadband acoustic cloaks using a core-shell configuration. The proposed model utilizes fully connected DNNs to capture and generalize the nonlinear intricate relation between the design parameters and the spectral response for the forward and the inverse problem [see Fig.1(a)]. The implementation of the forward problem is straightforward, and consists in training the neural network that maps the design parameters directly to the spectral response. Yet, the inverse design is intrinsically challenging due to non-uniqueness of the solution and inherent convergence problems[30,31]. To address these issues, we design an autoencoder-like structure consisting of two DNNs where the pretrained forward network is cascaded behind the inverse neural network that maps the spectral response to the design parameters in either a deterministic [See Fig.1(b)] or a probabilistic manner [See Fig.1(c)]. The deterministic inverse design network provides only one set of design parameters for a given spectral response. However, a practical implementation generally demands more flexibility and diversity in the design due to external perturbation. Hence, we introduce a novel model to provide probabilistic distributions of the design parameters, which flexibly generates the desired spectral response [see Fig.1(c)].

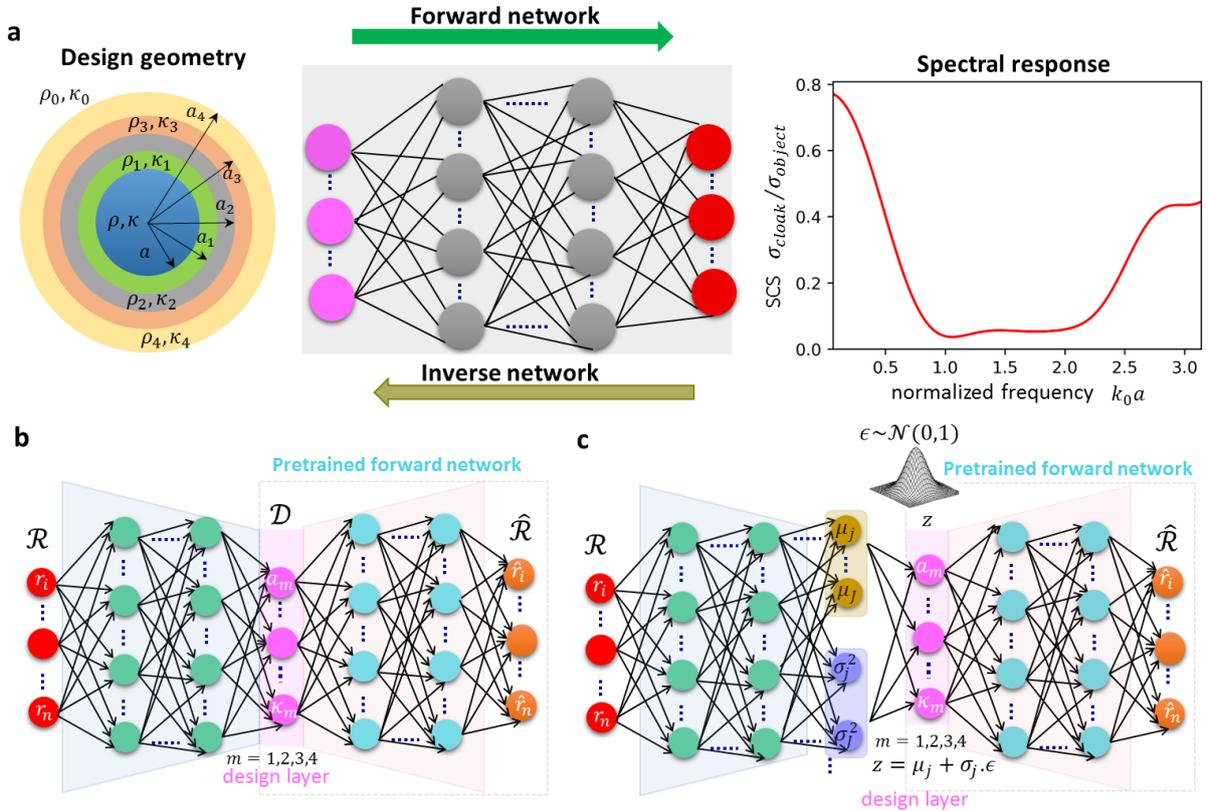

Fig. 1 Framework of the deep learning network for inverse design of the acoustic cloak. a Schematic illustration of the core-shell acoustic cloak and its spectral response (ratio between total scattering cross-section spectra of the cloak $\sigma_{\text{cloak}}$ and that of the object, i.e., $\sigma_{\text{object}}$) where the neural network learns the relation from $\mathcal{D}$ (design parameters) to $\mathcal{R}$ ( spectral response ) and from $\mathcal{R}$ to $\mathcal{D}$ for forward and inverse design, respectively. b-c Proposed

deep learning models for inverse design of the cloak. b Deterministic model where the pretrained forward network acts as decoder to predict the spectral response. c Probabilistic model where the design space is transformed into the latent space *z*, with a standard Gaussian distribution. The physical design parameters are sampled from that distribution in the form of latent variables to generate the desired spectral response.

The probabilistic design with parameter distributions is more advanced than the deterministic design with fixed parameters on the refection of twofold benefits (i) the capability to generating a variety of design parameters for one desired spectral response and (ii) the ability to uncovering the sensitivity of the design parameters on the cloaking effect. We find that the bulk modulus is less sensitive to external perturbation than the thickness of the layers in designing the acoustic cloaks. We also perform numerical simulations with the finite element method (FEM) to confirm the theoretical cloaking predictions.

**Results:**

**Deep learning model.** To study acoustic invisibility, we consider a four-layered shell configuration where each layer is parametrized with different material and outer radius as shown in Fig. 1a. The properties of the cylindrical scatterer are described by its radius $a$, volume mass density $\rho$, and bulk modulus $\kappa$, while the cloak's layers are sequentially numbered as $m=1,2,3,4$ to represent the outer radius $a_m$ and material properties $(\rho_m, \kappa_m)$ of the *m*th layer. In order to analyze the scattering response of this system, we make use of the transfer matrix method (TMM) and compute the total scattering cross-section (SCS) spectra [See Methods section for details]. To quantify the cloak's performance, we define the ratio of the SCS spectra of the cloaked object $\sigma_{\text{cloak}}$ and the bare object $\sigma_{\text{object}}$, i.e., $\sigma_{\text{cloak}}/\sigma_{\text{object}}$ (or normalized SCS). This ratio reveals how well the object becomes acoustically invisible with the presence of the designed cloak. The ideal cloaking behavior is achieved by optimizing the design parameters to yield a SCS as close to zero as possible at the operation frequency. In our study, we consider, without loss of generality, a cylindrical scatterer with parameters: $a = 1$ m, $\rho = \rho_0$, and $\kappa = 1.5 \times \kappa_0$, with $\rho_0$ and $\kappa_0$ being the mass density and the bulk modulus of the host medium, respectively. We use the TMM to generate the training data samples, where we determine the SCS spectra for random design space $\mathcal{D}=[a_1, a_2, a_3, a_4, \kappa_1, \kappa_2, \kappa_3, \kappa_4]$ with $a_m$ and $\kappa_m$ the radius and bulk modulus of the *m*th layer, respectively, while keeping the density fixed in each layer . Each training example is represented by eight design parameters (four radii and four bulk moduli) and 100 discrete points of SCS spectrum $\mathcal{R} = [r_1, r_2, r_3, r_4 \ldots \ldots \ldots r_{100}]$ covering the normalized frequency range $0 \leq k_0 a \leq \pi$, with $k_0 = 2\pi/\lambda$ the acoustic wavenumber. We design the forward neural network to map the design $\mathcal{D}$ to the spectrum $\mathcal{R}$, and the inverse model to map the spectrum $\mathcal{R}$ to the design $\mathcal{D}$. Both networks are trained by optimizing the neural network weights. For our analysis, we generate 68000 data samples for random design parameters, which are split into three distinct groups: 60000 data samples for training, 4000 data samples for validation, and 4000 data samples for final testing. The training data is used to train the network by optimizing the neural network weights, while the validation data set serves for checking and avoiding the overfitting issue, and the testing data set examines the prediction accuracy of the network.

**Forward-modeling network.** We first design the forward-modeling network to accurately predict the frequency-dependent SCS for given design parameters. The forward model builds a fully connected network between the design space $\mathcal{D}$ as the input layer and SCS spectra $\mathcal{R}$ as the output layer, as shown in Fig. 1(a). We normalize the data before training, to expedite the convergence of the network. In the training process, the training data is fed into the network and the weights are continuously optimized to minimize the loss function defined as $\mathcal{L} = \frac{1}{N}\sum_k |r_k - \hat{r}_k|$ where $r_k$ and $\hat{r}_k$ are the ground truth of the spectral response and the response predicted by the neural network, respectively. Note that we used the mean absolute error as a cost function for the training purpose due to the presence of outliers in the data. The architecture of the forward network is optimized to have four fully connected layers with each layer having 500–500–500–300 nodes, respectively. The remaining hyper parameters (batch size, learning rate, activation function, etc.) are judiciously selected to minimize the validation loss [See Supplemental Material (SM)]. The learning curves for the training and validation data as a function of epoch are shown in Fig. 2(a). The training and validation errors drop both as the training goes on, until it converges after 400 epochs, implying the completion of the training phase. Due to the unique one to one design-to-response mapping in the forward problem, the training process is straightforward to converge. To check the prediction accuracy of the trained model, we evaluate it on a group of 4000 data samples that are not seen in the training process. We predict the spectral response of the trained network for testing data and compare the results with those obtained by the TMM. We report the relative absolute spectral error on the testing data sets as: $e = \sum_k |r_k - \hat{r}_k|/r_k$ where $r_k$ is the discretized value for the target spectral response and $\hat{r}_k$ is the corresponding predicted spectral response. The results for relative spectral error are plotted in Fig. 2(b) with a mean error of 3.47% for testing data, which proves the high prediction accuracy (over 96%) of the network. Figures (c)-(e) depict the spectral response of three representative cases from the testing data that clearly indicate the predicted results perfectly match with those of the TMM.

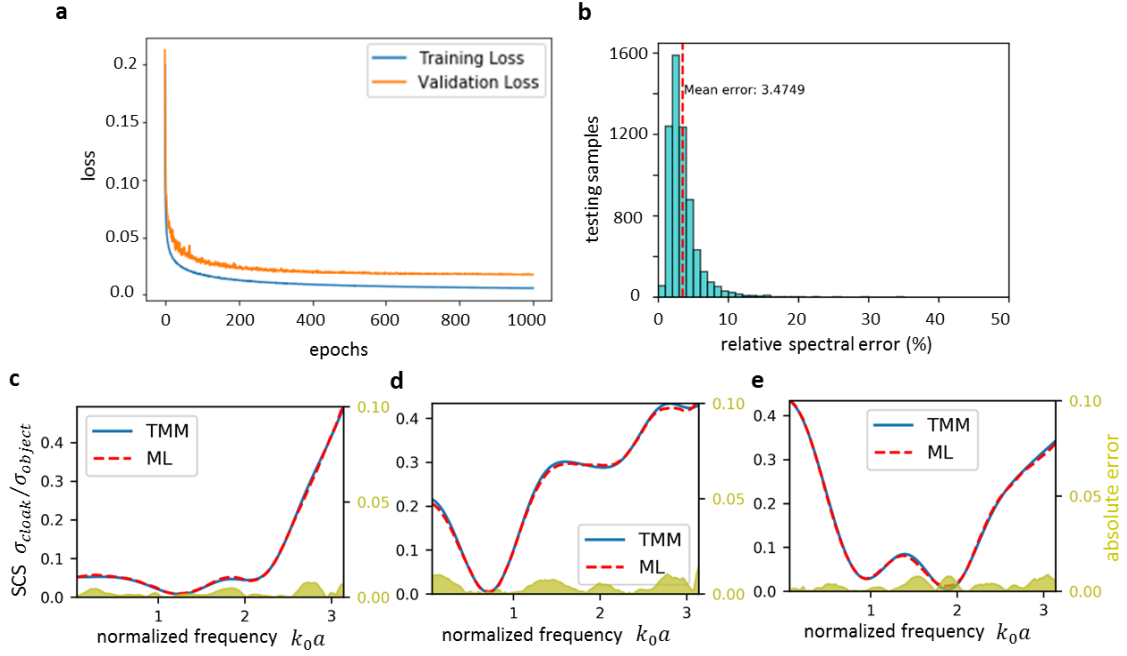

Fig. 2 Forward network learning of acoustic cloak. a Learning curves for training and validation datasets as a function of training epochs. b Histogram of relative spectral error for testing samples. The red vertical dashed line shows the mean spectral error. c-f Comparison of the spectral response for three representative examples obtained by machine learning model and TMM. The shaded area shows the absolute error between the predicted and target response which indicates that the machine learning results are in perfect agreement with those of TMM. The design parameters are provided in SM.

**Deterministic inverse-modeling network**. Typically, practical applications demand cloaking effect at a particular frequency or broad frequency band, which necessitates the experimentally realizable design parameters to produce the desired spectral response. Yet, there is no common tool available for the accurate inverse design of clocking devices. The development of such tools significantly reduces the computational time for design optimization and accelerates the generation of the desired SCS spectra. To achieve this goal, we attempt to train the network inversely, which takes the spectral response as the input and the design parameters as the output. However, we are unable to train the network successfully in the inverse direction due to non-unique solutions in response-to-design mapping that leads to non-convergence issues [See the SI for more details]. To resolve this issue, we implement the autoencoder-like network where we cascade the inverse network to the independently trained (or pretrained) forward-modeling network, as shown in Fig. 1(b). The forward network is trained separately to substitute the TMM simulation and acts as a data generator in training. During the training process, the pretrained forward network has fixed weights and biases while the weights of the inverse network are updated iteratively to minimize the loss function to the output response of designed structure predicted by neural network. The designed structure refers to the intermediate layer $\mathcal{D}$ in the autoencoder-like network that predicts the eight-design parameter for the desired spectral response.

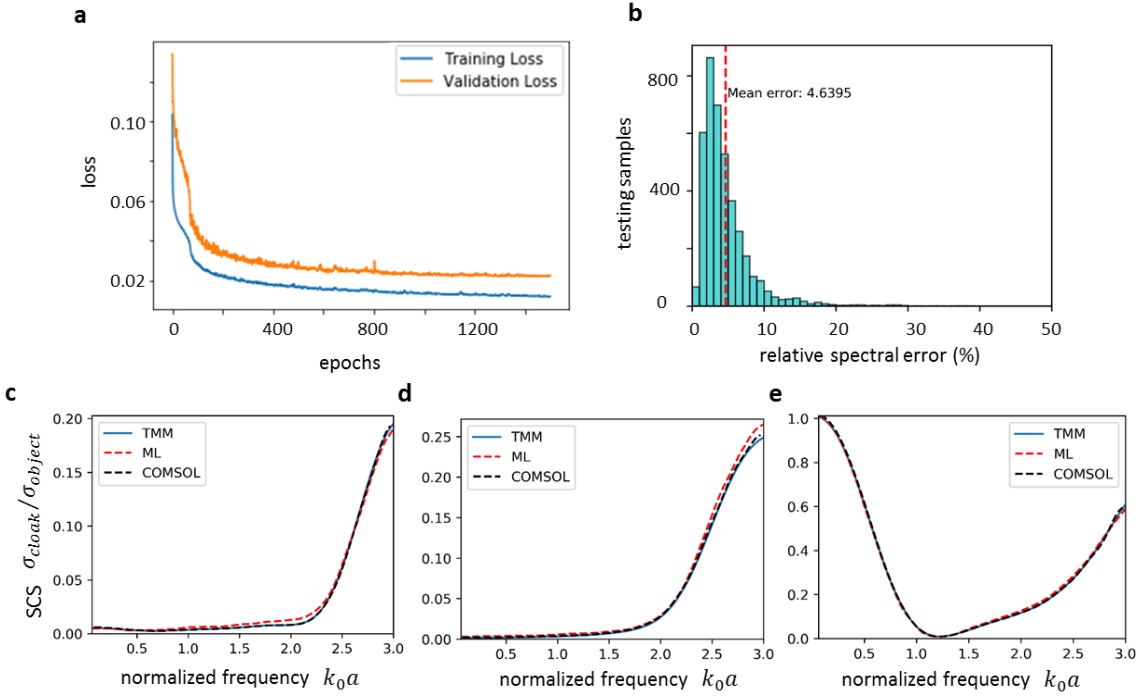

Fig. 3 Deterministic inverse design of acoustic cloak. a Learning curves for training and validation datasets as functions of the training epochs. b Histogram of the relative spectral error for the testing data samples. The red vertical dashed line shows the mean spectral error. c-e Comparison of the spectral response for three representative cases obtained with machine learning, TMM, and COMSOL. These results clearly show that machine learning accurately predicts the target response. The designed responses require positive mass density and bulk modulus values that are provided in the SM.

Figure 3(a) depicts the learning curves for the training and validation loss that decrease rapidly and converge after 1000 epochs of training. The inverse network architecture is designed to have five fully connected layers with each layer composed of 500–500–500–500–400 nodes, respectively and details about the hyper parameters are discussed in the SI. Again, we use 4000 testing data sets to examine the accuracy of the inverse design network. Figure 3(b) shows the histogram of relative spectral error, where the average spectral error across all predicted spectra is 4.63%. The scattering response for representative cases (d)-(e) confirms the accuracy of designed inverse network.

Next, we perform full-wave numerical simulations to verify the predicted cloaking effect in the broad spectral range. Figure 4 depicts the real part of the pressure field distributions of the bare object and the cloaked object under the excitation of a plane wave with different frequencies. The results are obtained by finite element method (COMSOL Multiphysics). It shows the incident waves are severely distorted in the case of the bare object, indicating strong scattering behavior. However, the field distributions with the presence of the cloaked object are similar to those without any object, as if the cloaked object does not exist, validating the performance of the designed cloak, in a broad spectral range.

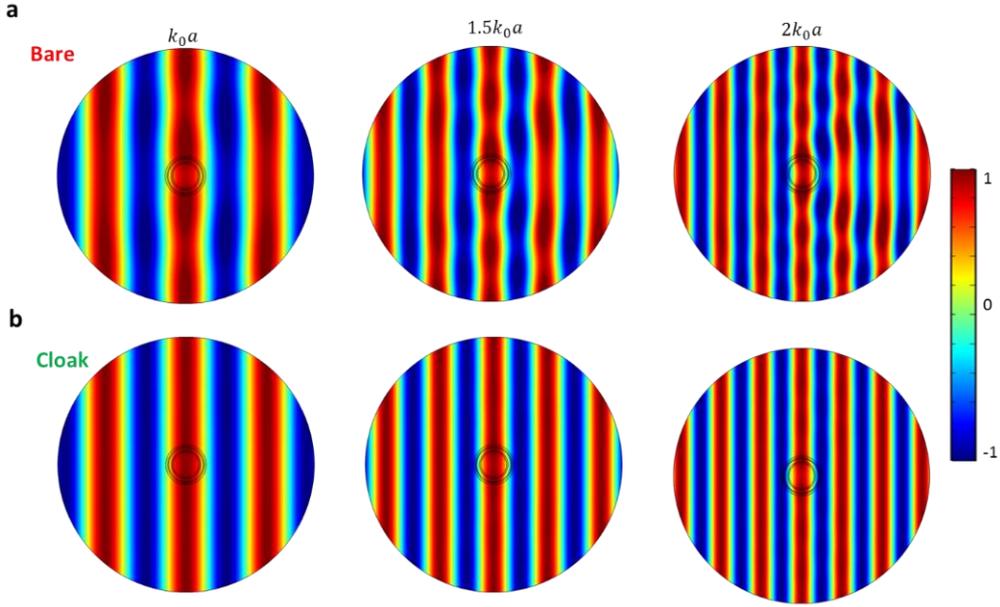

Fig. 4 Numerical simulations for designed broadband acoustic cloak. a Pressure field distribution produced by an incident plane wave for the bare object and b the cloaked object. The three different normalized frequencies $k_0 a = 0.5, 1, 1.5$ are picked from the spectral response shown in Fig. 2(c).

**Probabilistic inverse-modeling network.** In the deterministic inverse design, we choose one precise set of design parameters to generate the desired spectral response, yet, the practical implementation demands diversity in design parameters due to the possible unavailability of actual materials and unexpected deviations in the original design parameters. In this scenario, it is essential to enhance the generalization and robustness of our network by introducing the probabilistic prediction. To achieve this goal, we propose the stochastic inverse design that uses the latent space concept[34] for the probabilistic representation of the physical design parameters. Our probabilistic inverse design network is basically a generative model which transforms the input spectral response into a mean vector $\mu$ and a variance vector $\sigma^2$ to approximate the distribution of the latent variables corresponding to the design space, and then the pretrained forward network, acting as a decoder, generates the same spectral response as the input by sampling latent variable vector $z$ from the Gaussian prior distribution illustrated in Fig.1(c). The loss function for the probabilistic inverse network consists of mean absolute error for reconstruction $\mathcal{L}_{MAE}$, a Kullback-Leibler (KL) divergence error $\mathcal{L}_{KL}$, and a precision parameter, i.e., the inverse of the variance $\tau_p = 1/\sigma^2$. The KL divergence ensures the generated latent space distributions follow the assumed Gaussian distribution, $\mathcal{N}(0,1)$, and also provides the discrepancy between the predicted distributions and the standard normal distribution. The precision term is incorporated to avoid the zero-variance problem for the generated distributions of the latent space. The model is trained to minimize the following loss function:

$$\mathcal{L} = \frac{1}{N}\sum_{k=1}^{N}\left(\mathcal{L}_{MAE}^{(k)} + \alpha\mathcal{L}_{KL}^{(k)} + \beta\tau_p^{(k)}\right), \qquad (1)$$

where $N$ is the total number of training samples, $\alpha$ is the weight of probabilistic learning, and $\beta$ is the regularization parameter. These hyper parameters can be tuned by cross-validation during the training process. The detail of the derived cost function is provided in the Methods section.

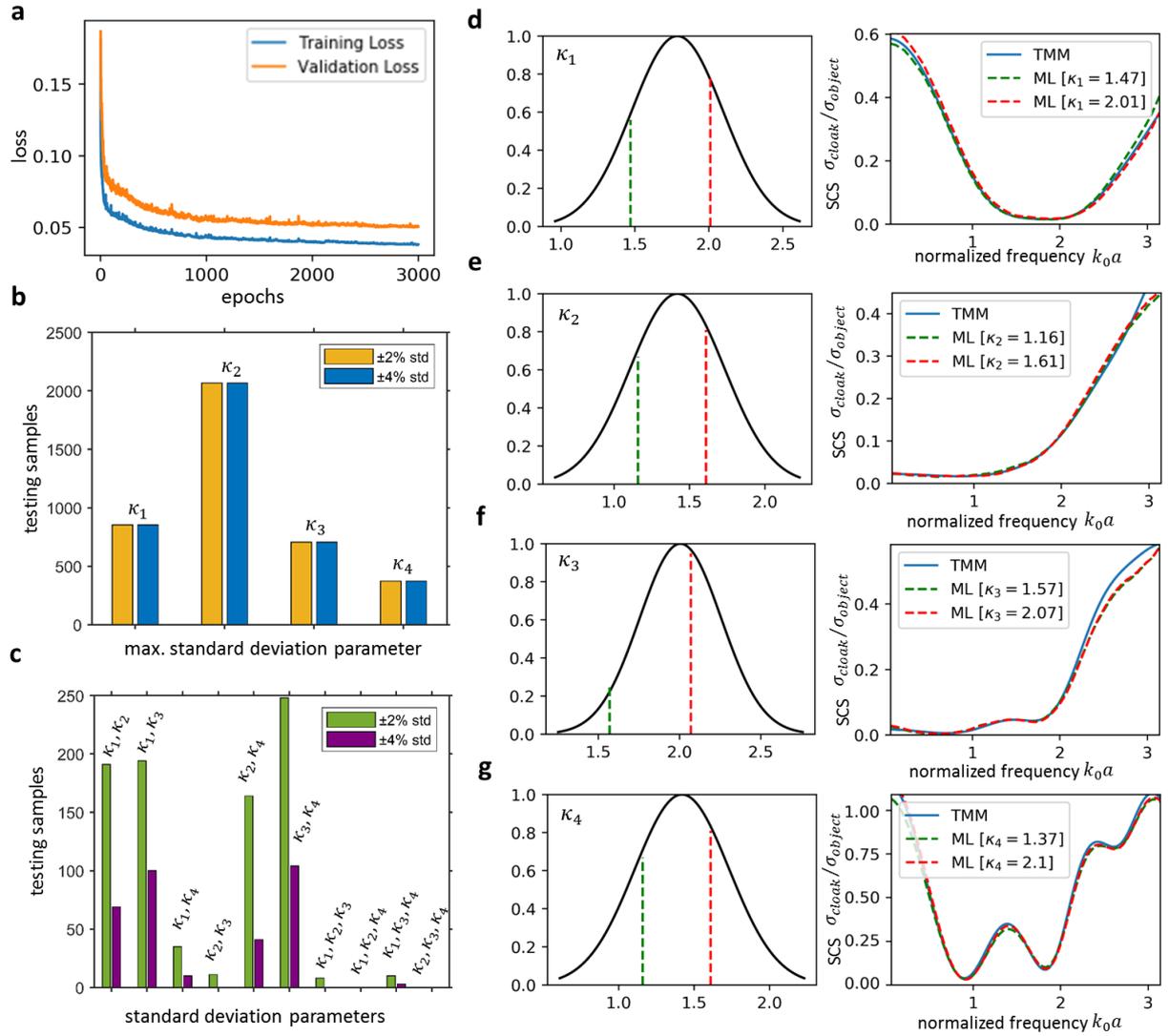

Fig. 5 Probabilistic inverse design of acoustic cloak. a Learning curves for the training and validation datasets over training epoch. b Classification of testing samples based on maximum standard deviation (±2% and ±4% deviation

from the mean value) in the generated distributions of design parameters. Bulk modulus parameters in design space exhibit large standard deviation. c classification of testing samples based on ±2% and ±4% deviation from the mean value in the generated distributions of different combinations of bulk modulus parameters. d-g Gaussian distributions of bulk moduli showing maximum standard deviation for the representative cases. Comparison of the spectral responses generated from randomly sampled bulk moduli in the corresponding distribution (dashed red and green curves) and reference response (solid blue curve).

Figure 5(a) shows the training and validation learning curves as a function of the training epoch, where the loss function for both training and testing datasets converges after 2500 epochs. The architecture of stochastic inverse model is designed to have nine fully connected layers with each layer having 500–500–500–500–500–500–500–500–400 hidden neurons, respectively [see the details about the hyper parameters in the SI]. The designed network essentially learns the distribution of design parameters conditioned on its spectral response. After successful training, we explore the generated distributions of the design space for the testing data. The standard deviation of the generated distribution indicates the range of design parameters that can generate the same spectral response. It also reveals the sensitivity of the design parameters on the generation of the spectral response. Quantitatively, we can classify the spectral responses by finding the maximum standard deviation in the latent distributions of the design parameters for testing datasets. We observe that the distributions of the bulk modulus parameters exhibit large standard deviations as compared to the thicknesses in the design space. For the analysis, we set criteria of minimum ±2% and ±4% standard deviation (from the mean value of the corresponding design parameter) in the distribution and categorize the predicted design parameters in testing samples based on the maximum standard deviation, as depicted in Fig. 5(b). We further sort the design space in testing data based on minimum ±2% and ±4% standard deviation (with respect to mean value) in different combination of bulk modulus parameters as shown in Fig. 5(c). Such classifications identify how the different sets of the bulk moduli may be used to generate the same spectral response. Figures 5(d)-(g) show the spectral response of four representative cases with maximum standard deviation in the bulk modulus distributions. The randomly sampled bulk moduli in the corresponding distribution (indicated by dashed red and green curves) generate the same spectral responses, showing the excellent performance of our stochastic design [See more examples in the SM].

**Discussion**

We use supervised and semi-supervised learning algorithms based on the nonlinear regression technique to develop the deep learning model that maps the design space to the spectral response for the forward design and the spectral response to the design space in the inverse design of the acoustic cloak. The mapping functions are implemented by deep learning neural networks. To solve the inverse problem, we design the deterministic and probabilistic encoder-decoder like networks. The deterministic design encodes the spectral response into the design parameters, which are decoded with the pretrained forward model to generate the desired response. Yet, the probabilistic design models the statistical distribution of the design parameters. Such distributions allow the diversity and flexibility for fabrication and application of the designed cloak. The training of the probabilistic design process includes four basic steps: First, the spectral response is transformed into a distribution over the latent space. Second, a point from the latent space is randomly sampled from that distribution. Third, the sampled point is decoded using a pre-trained forward model (acting as a decoder) and the reconstruction error is computed. Fourth, the reconstruction error is back propagated through the network to minimize the loss function. We use the normalized parameters for operating frequency and material properties, which can be scaled depending on the choice of the host medium and scatterer. For example, for a scatterer of size of 10 cm immersed in water, the designed cloak perfectly works over a broad spectrum ranging from a few Hz to 5 kHz. To demonstrate this idea, we consider a specific core material to achieve broadband cloaking by tuning the material and geometry of the four layered core-shell system. However, the approach can be applied to design invisible cloaks for any given core in such systems.

To summarize, we demonstrate the machine learning driven broadband acoustic cloak with multilayer core–shell configuration. In particular, we develop deterministic and probabilistic deep learning models for inverse design of acoustic cloak that efficiently solves the inverse design problem. The proposed models utilize the encoder-decoder like structure to solve the on-many mapping problem and retrieve the design parameters for the given spectral response. The forward network, acting as a decoder, is trained independently and cascaded behind the inverse network either in deterministic or probabilistic design. In the probabilistic design, the design parameters in the form of probabilistic latent variables, are obtained by the sampling from distributions in the latent space. The distributions of the design parameters are used to reveal the sensitivity of the design parameter on the cloaking functionality and the generation of the desired spectral response with diverse design. The probabilistic network is highly attractive route to improve the robustness of the cloaking effect against the deviation of the design parameters of the cloak. We envision that our approach can be generically utilized to automate the designing process of complex material systems, showing non-unique solution space, with minimum human intervention.

## Methods

**Transfer Matrix method**. The TMM is used to compute the scattering response of the considered acoustic core-shell cloak due to the cylindrical invariance of the problem. The incident and scattered pressure fields are expanded using Bessel and/or Hankel functions in all regions. The pressure field in the host medium $P^{(0)}$ is the summation of the incident field $P_{inc}^{(0)}$ and the scattered one $P_{sca}^{(0)}$, i.e.,

$$P^{(0)}(r) = \sum_{n=0}^{\infty} \varepsilon_n i^n [J_n(k_0 r) + \zeta_n^{sca} H_n^1(k_0 r)] \cos(n\theta), \qquad (2)$$

where $k_0$ is the wavenumber in the host medium, $\zeta_n^{sca}$ represents the complex scattering coefficient, $J_n$ is the Bessel function of order $n$ and $H_n^1$ is the Hankel function of the first kind and of order $n$. For convenience, we introduce the parameter $\varepsilon_n$ being 1 for $n = 0$ and 2 for $n \geq 1$, that stretch the summation index from one to infinity, but needs to be truncated for convergence.

The pressure field in each cloaking layer is expressed using both Bessel functions, as there is no singularity (i.e., $r \neq 0$)

$$P^{(m)}(r) = \sum_{n=0}^{\infty} \varepsilon_n i^n \left[ \zeta_n^{(m)} J_n(k_m r) + \chi_n^{(m)} Y_n(k_m r) \right] \cos(n\theta), \qquad (3)$$

where the superscripts $m = 1,2,3,4$ refer to the cloaking layers, $\zeta_n^{(m)}$ and $\chi_n^{(m)}$ are the unknown coefficients in the corresponding cloaking layer. On the contrary, the pressure field inside the scatterer can be expressed only using the Bessel functions of the first kind, i.e.,

$$P(r) = \sum_{n=0}^{\infty} \varepsilon_n i^n \zeta_n J_n(k_1 r) \cos(n\theta). \qquad (4)$$

The coefficients $\zeta_n$, $\zeta_n^{(m)}$, $\zeta_n^{sca}$, $\chi_n^{(m)}$ are needed to determine the scattering response of the scatterer. By applying the boundary condition, i.e., the continuity of pressure $P$ and the normal component of the velocity (proportional to $(1/\rho)\partial_r P$) across all interfaces, we can obtain the $T$-matrix that relates the incident and scattered fields. The scattering coefficient $\zeta_n^{scat} = T_{11}/T_{21}$ computed from the components of $T$-matrix is used to determine the SCS that estimates the total power scattered by the cloaked cylinder, and is hence a measure of its far-field 'visibility', given as,

$$\sigma_{\text{cloak}} = \frac{4}{k_0} \sum_{n=0}^{N} |\zeta_n^{\text{sca}}|^2 . \tag{5}$$

To cloak the object, we need to minimize the SCS by searching the suitable bulk modulus and thickness of the cloaking layers. In the data generation process, the spectral response computed from $\sigma_{\text{cloak}}/\sigma_{\text{object}}$ is the anticipated output. We apply this approach for data generation due to higher computational efficiency and accuracy as compared to finite element and finite difference time domain numerical methods.

**Probabilistic design**. The probabilistic inverse model provides the distribution of the design parameters, given the spectral response and the latent variable $z$. This model uses the variational inference approach to approximate the distribution from which the latent variable is drawn. Therefore, the loss function is modified in comparison to the deterministic case. The loss function contains the KL divergence term that guarantees our learned distribution $q(z|x)$ to be analogous to the predefined distribution $p(z)$, which is assumed to be Gaussian, for each dimension of the latent space. In addition, we include the precision parameter to avoid the zero variance issue arose in biased data. The modified loss function for the stochastic network is defined as

$$\mathcal{L}^{(k)} = \mathcal{L}_{MAE}^{(k)} + \alpha \sum_{j=1}^{8} \mathcal{L}_{KL}^{(k)}\big(q_j(z|x) \parallel p(z)\big) + \beta \sum_{j=1}^{8} \tau_p^{(j)}, \tag{6}$$

where the superscript $k$ donates the $kth$ training sample, $j$ is the dimensionality of the latent space $z$, $\alpha$ is the weight of loss on KL-divergence, and $\beta$ is the regularization parameter. For the Gaussian prior distribution $\mathcal{N}(0,1)$, the precision parameter is the inverse of the variance $\tau_p^{(j)} = 1/\sigma_j^2$ and the KL divergence between two Gaussian distributions can be expressed as

$$\mathcal{L}_{KL}^{(i)} = -\frac{1}{2} \sum_{j=1}^{8} 1 + \sigma_j^2 - \mu_j^2 - log(\sigma_j), \tag{7}$$

where $\mu$ and $\sigma^2$ are the mean and the variance of the generated latent space distribution, respectively. During the optimization process, reparameterization operation is used to compute $z = \mu_j + \sigma_j \varepsilon$, where $\varepsilon$ is a sample from the standard Gaussian distribution. This procedure allows for using the gradient-based optimization method by moving the sampling to an input layer. The deep learning models are implemented in Python using open-source neural-network library Keras and tensor flow.

**Numerical modeling.** The full wave simulations are performed with the acoustic module in COMSOL Multiphysics based on the finite element method. In these simulations, water (with mass density $\rho = 998$ kg/m$^3$, bulk modulus $\kappa = 2.91$ GPa) is used as the host medium. The core-shell structure is excited with an acoustic plane wave (of unit-amplitude) to determine the scattering response of the cylindrical object.

## Acknowledgements

The work described in here is partially supported by King Abdullah University of Science and Technology (KAUST) Office of Sponsored Research (OSR) under Award No. OSR-2016-CRG5-2950 and KAUST Baseline Research Fund BAS/1/1626-01-01.

# Supplemental Material

In this supplementary material, we discuss non-convergence and overfitting issue in the inverse problem, details about optimized hyper parameters and presents more examples for the designed deterministic and stochastic inverse networks.

## 1.1 Overfitting problem in designing direct inverse network

As discussed in the main manuscript, the inverse design involves training of the network to retrieve the design parameters from the spectral response. Nevertheless, the direct training of inverse network suffers from overfitting problem that is caused due to one-to-many possible mapping and, as a result, the inverse function does not converge. Figure S1 depicts the training and validation losses during learning process where the training loss decreases continuously but the validation loss initially decreases and then increases to grasp the convergence. Such trend shows the overfitting behavior and does not reveal the mapping present in the data that leads to inaccurate predictions. To overcome such issues, we design encoder-decoder like network to determine one possible solution (and eliminate the others) deterministically and stochastic network to find different sets of possible solutions. The designed deep learning models can be used to predict the scattering response for any (given) design parameters and vice versa.

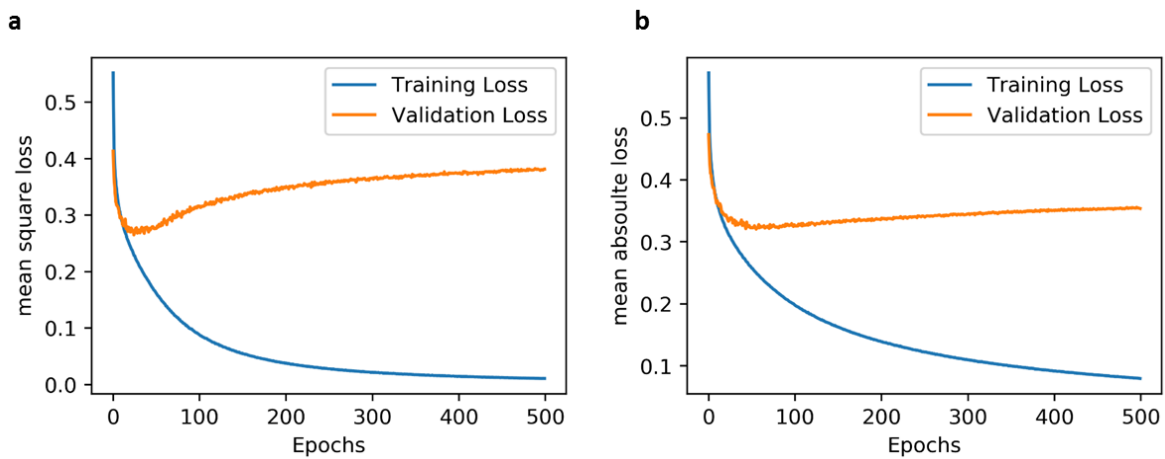

Fig. S1. The loss function of training and validation errors during learning process of a direct inverse network. (a) Mean square loss and (b) mean absolute loss. In both cases, the validation error initially decreases until a point at which it begins overfitting that indicate the network's poor performance in designing parameters for the given spectral response. The architecture is composed of 100-500-500-500-400-8 nodes.

## 1.2 Optimal setting of hyper parameters for the designed forward and inverse network

The training results of the neural network is largely determined by the structure of the network and every hyper parameter. After various tests with the same training data samples, the hyper parameters can be fixed for the forward and inverse networks, which are shown in Table S1.

Table S1. Hyperparamters for the optimized networks

| Parameters | Forward network | Inverse network (deterministic) | Inverse network (probabilistic) |
|---|---|---|---|
| Activation function | ReLu | LeakyReLu f(x)= (0,0.2max) | LeakyReLu f(x)= (0,0.2max) |
| Optimizer | Adam (learning rate=0.0005, decay rate=1e-5) | Adam (learning rate=0.0002, decay rate=1e-5) | Adam (learning rate=0.0002, decay rate=1e-5) |
| Batch size | 100 | 100 | 100 |
| Hidden layers (neurons) | [500-500-500-300] | [500-500-500-500-400] | [500-500-500-500-500-500-500-400] |

## 1.3 Design parameters for the cases considered in the main text

The design parameters of the results discussed in the main text are summarized in Table S2. In all cases, the densities of the designed layers are $\rho$ = [0.308, 1.961, 1.557, 0.579]. Note that we provide the normalized parameters which can be easily transformed to realistic parameters depending the on the choice of the host medium.

Table S2. Design parameters for the predicted response in the main text

| case | $a_1$ | $a_2$ | $a_3$ | $a_4$ | $\kappa_1$ | $\kappa_2$ | $\kappa_3$ | $\kappa_4$ |
|---|---|---|---|---|---|---|---|---|
| Fig.2c | 1.110 | 1.441 | 1.544 | 1.756 | 1.921 | 0.645 | 1.838 | 0.937 |
| Fig.2d | 1.093 | 1.301 | 1.382 | 1.577 | 1.375 | 0.707 | 0.517 | 0.851 |
| Fig.2e | 1.110 | 1.388 | 1.515 | 1.623 | 1.272 | 0.630 | 1.304 | 1.605 |
| Fig.3c | 1.105 | 1.347 | 1.417 | 1.567 | 0.556 | 0.887 | 0.823 | 0.913 |
| Fig.3d | 1.097 | 1.319 | 1.442 | 1.594 | 0.581 | 0.750 | 0.922 | 1.016 |
| Fig.3e | 1.228 | 1.404 | 1.61 | 1.788 | 0.750 | 1.019 | 0.539 | 1.287 |
| Fig.5d | 1.01 | 1.141 | 1.351 | 1.504 | 1.786 | 0.506 | 1.114 | 1.417 |
| Fig.5e | 1.152 | 1.154 | 1.639 | 1.773 | 0.950 | 1.617 | 0.774 | 1.176 |
| Fig.5f | 1.066 | 1.321 | 1.337 | 1.462 | 0.267 | 1.095 | 2.005 | 1.004 |
| Fig.5g | 1.067 | 1.203 | 1.491 | 1.519 | 0.534 | 0.468 | 1.890 | 1.898 |

**1.4 Additional examples for forward and inverse design network**

We present some additional results from the testing data to show the accuracy of our designed forward and inverse networks.

**I. Forward design:**

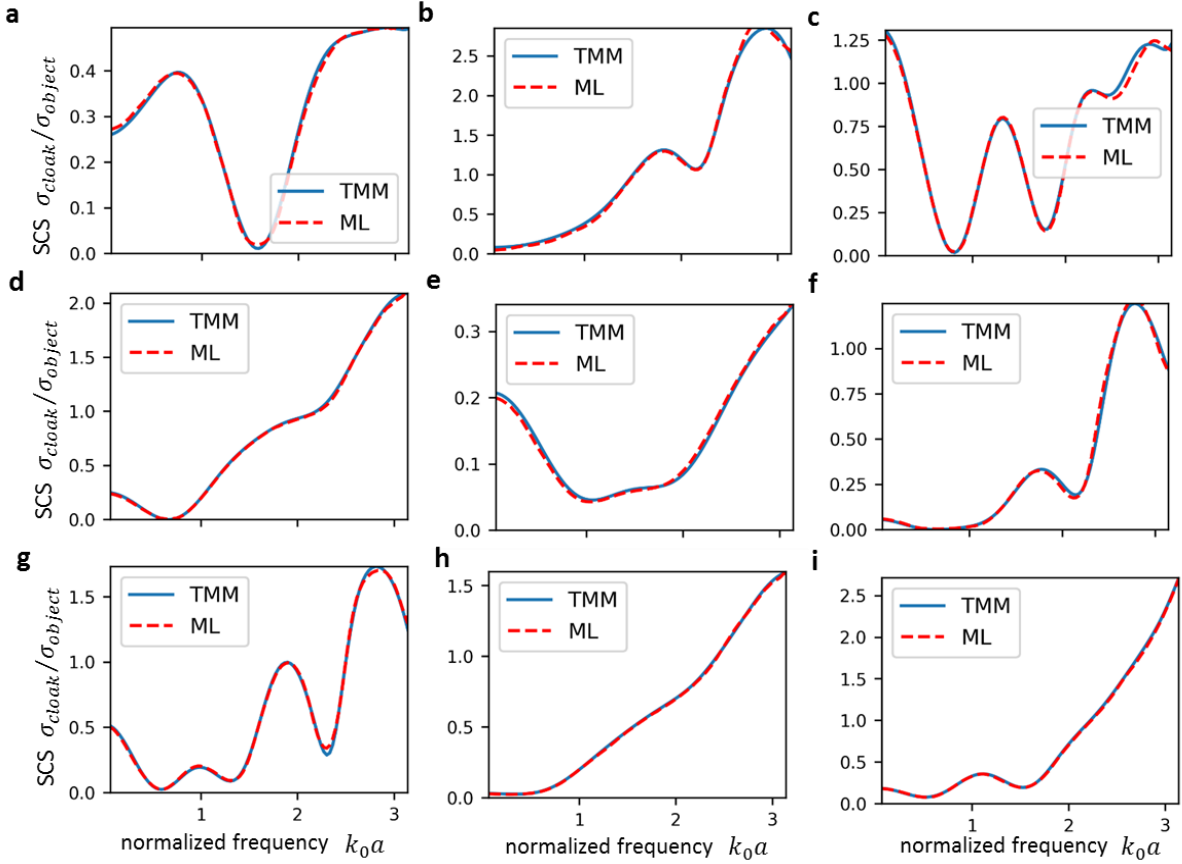

Fig. S2. Additional examples for forward design network. The parameters are provided in the Table S3.

Table S3. Design parameters for the predicted response in the Fig. S3

| case | $a_1$ | $a_2$ | $a_3$ | $a_4$ | $\kappa_1$ | $\kappa_2$ | $\kappa_3$ | $\kappa_4$ |
|---|---|---|---|---|---|---|---|---|
| Fig. S2a | 1.255 | 1.509 | 1.650 | 1.745 | 0.922 | 0.592 | 0.850 | 1.895 |
| Fig. S2b | 1.270 | 1.313 | 1.853 | 1.953 | 1.057 | 0.3606 | 1.448 | 0.3741 |
| Fig. S2c | 1.222 | 1.419 | 1.787 | 1.958 | 1.253 | 0.622 | 1.766 | 0.820 |
| Fig. S2d | 1.138 | 1.329 | 1.40 | 1.520 | 1.554 | 1.468 | 0.225 | 0.866 |
| Fig. S2e | 1.117 | 1.361 | 1.433 | 1.482 | 1.770 | 0.571 | 1.690 | 1.636 |
| Fig. S2f | 1.178 | 1.309 | 1.694 | 1.860 | 0.602 | 1.447 | 1.1089 | 0.621 |
| Fig. S2g | 1.161 | 1.529 | 1.634 | 1.963 | 0.782 | 1.081 | 1.542 | 0.671 |
| Fig. S2h | 1.167 | 1.321 | 1.391 | 1.444 | 1.637 | 0.989 | 0.274 | 1.211 |
| Fig. S2i | 1.174 | 1.148 | 1.649 | 1.727 | 1.952 | 0.501 | 1.767 | 1.738 |

## II. Deterministic inverse design:

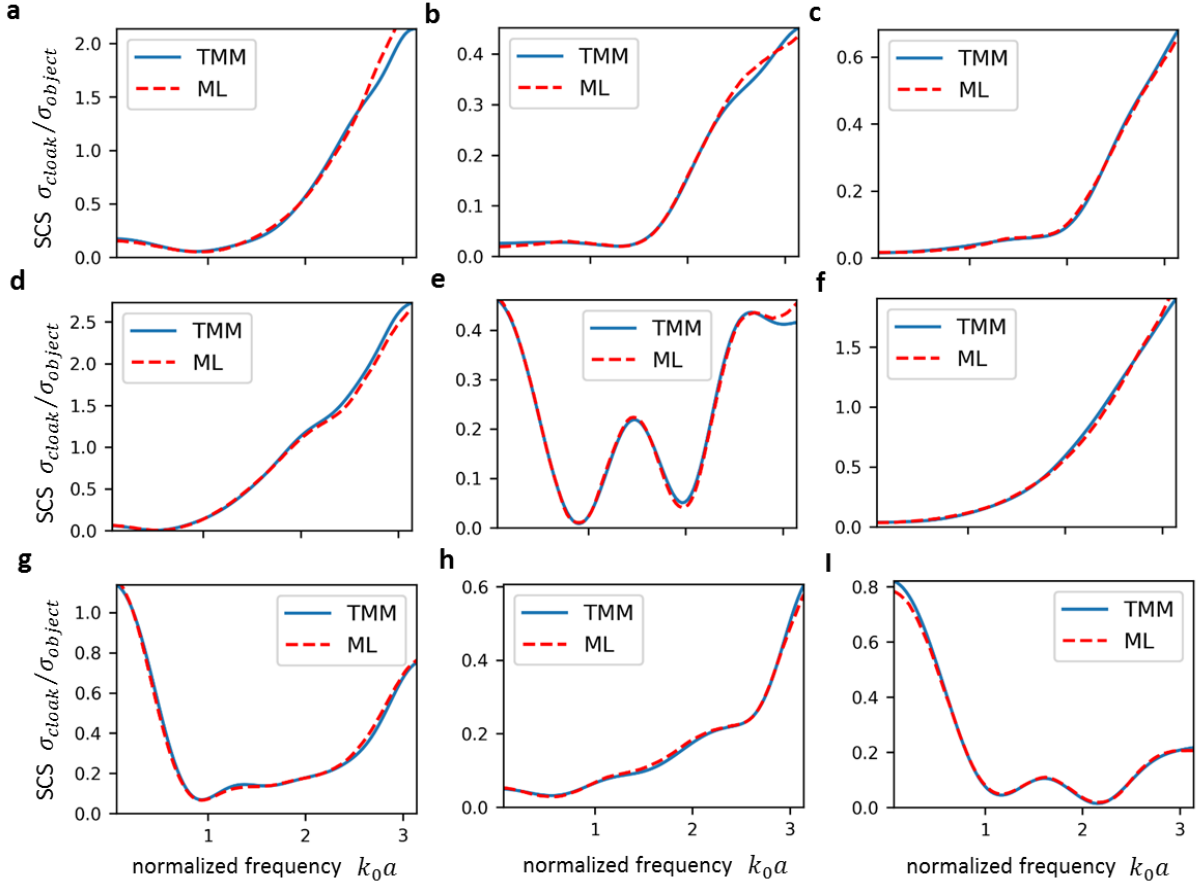

Fig. S3. Additional examples for deterministic inverse design network. The parameters are provided in the Table S4.

Table S4. Design parameters for the predicted response in the Fig.S3

| case | $a_1$ | $a_2$ | $a_3$ | $a_4$ | $\kappa_1$ | $\kappa_2$ | $\kappa_3$ | $\kappa_4$ |
|---|---|---|---|---|---|---|---|---|
| Fig.S3a | 1.247 | 1.327 | 1.835 | 2.006 | 1.427 | 1.106 | 0.675 | 1.437 |
| Fig. S3b | 1.163 | 1.180 | 1.590 | 1.670 | 0.871 | 0.814 | 0.745 | 1.463 |
| Fig. S3c | 1.178 | 1.251 | 1.721 | 1.897 | 0.993 | 0.860 | 0.788 | 1.197 |
| Fig. S3d | 1.229 | 1.257 | 1.5399 | 1.611 | 0.635 | 1.010 | 0.709 | 1.298 |
| Fig. S3e | 1.100 | 1.188 | 1.521 | 1.590 | 0.505 | 0.475 | 1.314 | 0.978 |
| Fig. S3f | 1.199 | 1.377 | 1.722 | 1.869 | 1.434 | 1.108 | 0.629 | 1.504 |
| Fig. S3g | 1.1859 | 1.327 | 1.632 | 1.910 | 0.891 | 0.911 | 0.630 | 1.049 |
| Fig. S3h | 1.154 | 1.342 | 1.567 | 1.737 | 0.880 | 1.045 | 0.686 | 0.974 |
| Fig. S3i | 1.067 | 1.358 | 1.446 | 1.542 | 0.796 | 0.800 | 1.083 | 1.588 |

## III. Probabilistic inverse design:

In probabilistic design, the Gaussian distributions for the design parameters are determined for a given spectral response. As we showed in the main text, the spectral response are categorized on the basis of the standard deviation in generated distribution of designs parameters. Here, we show some examples where different sets of bulk modulus are used to generate the same spectral response.

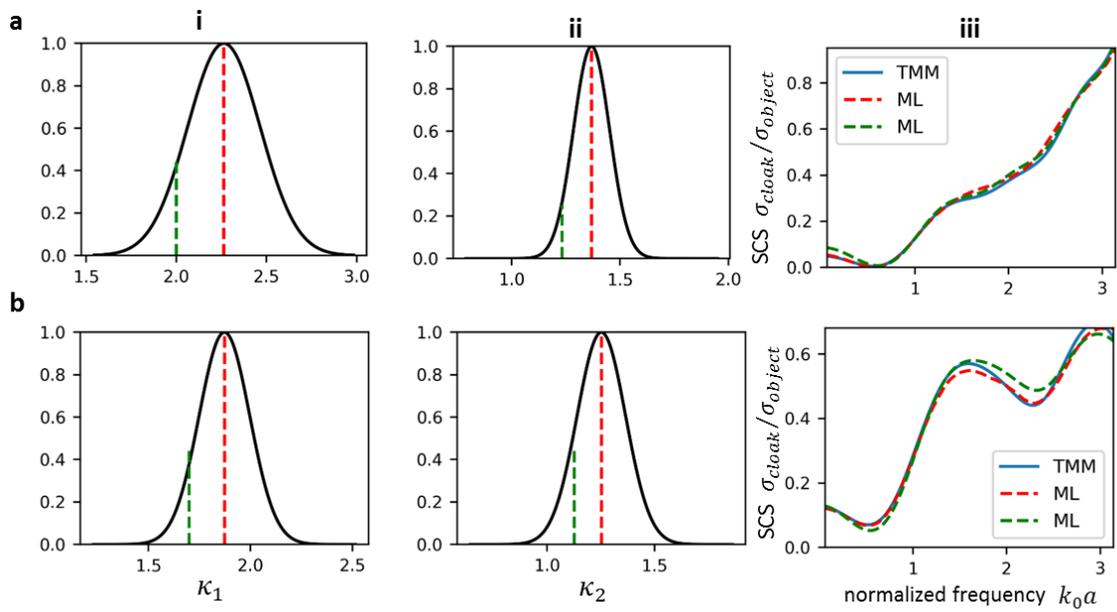

Fig. S4. Examples for probabilistic inverse design network showing maximum deviation in generated distributions of $\kappa_1$ and $\kappa_2$. The parameters are provided in the Table S5.

Table S5. Design parameters for the predicted response in the Fig.S4

| case | $a_1$ | $a_2$ | $a_3$ | $a_4$ | $\kappa_1$ | $\kappa_2$ | $\kappa_3$ | $\kappa_4$ |
|---|---|---|---|---|---|---|---|---|
| Fig. S4a | 1.054 | 1.094 | 1.464 | 1.698 | 2.263 | 1.369 | 0.596 | 1.285 |
| Fig. S4b | 1.055 | 1.083 | 1.335 | 1.535 | 1.872 | 1.253 | 0.771 | 0.714 |

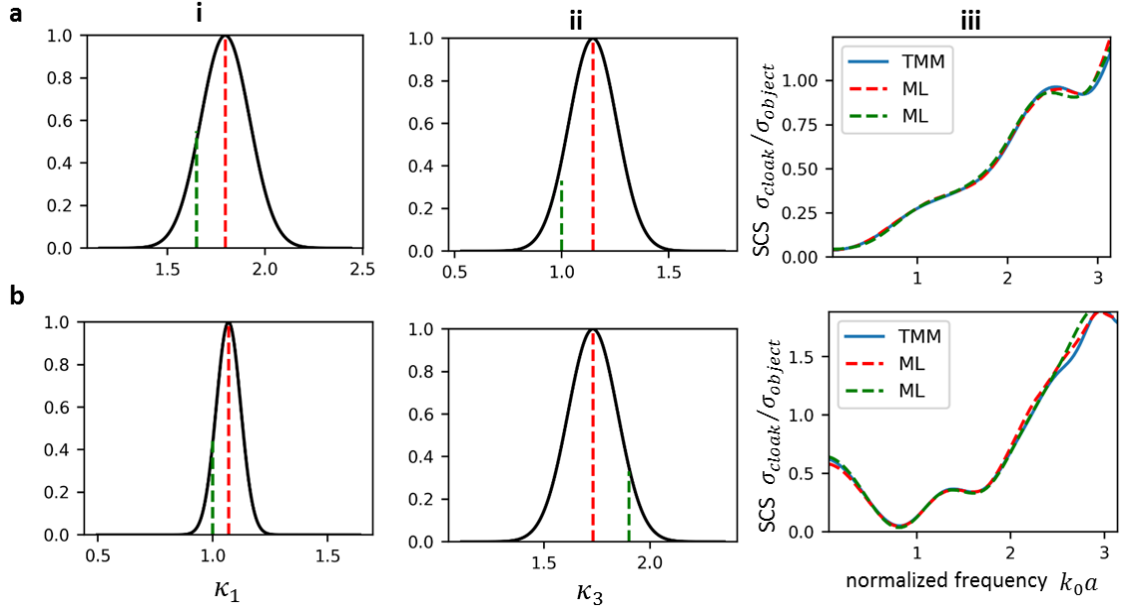

Fig. S5. Examples for probabilistic inverse design network showing maximum deviation in generated distributions of $\kappa_1$ and $\kappa_3$. The parameters are provided in the Table S6.

Table S6. Design parameters for the predicted response in the Fig.S5

| case | $a_1$ | $a_2$ | $a_3$ | $a_4$ | $\kappa_1$ | $\kappa_2$ | $\kappa_3$ | $\kappa_4$ |
|---|---|---|---|---|---|---|---|---|
| Fig. S5a | 1.122 | 1.46 | 1.481 | 1.606 | 1.796 | 0.892 | 1.145 | 0.521 |
| Fig. S5b | 1.05 | 1.429 | 1.559 | 1.772 | 1.069 | 0.64 | 1.731 | 1.407 |

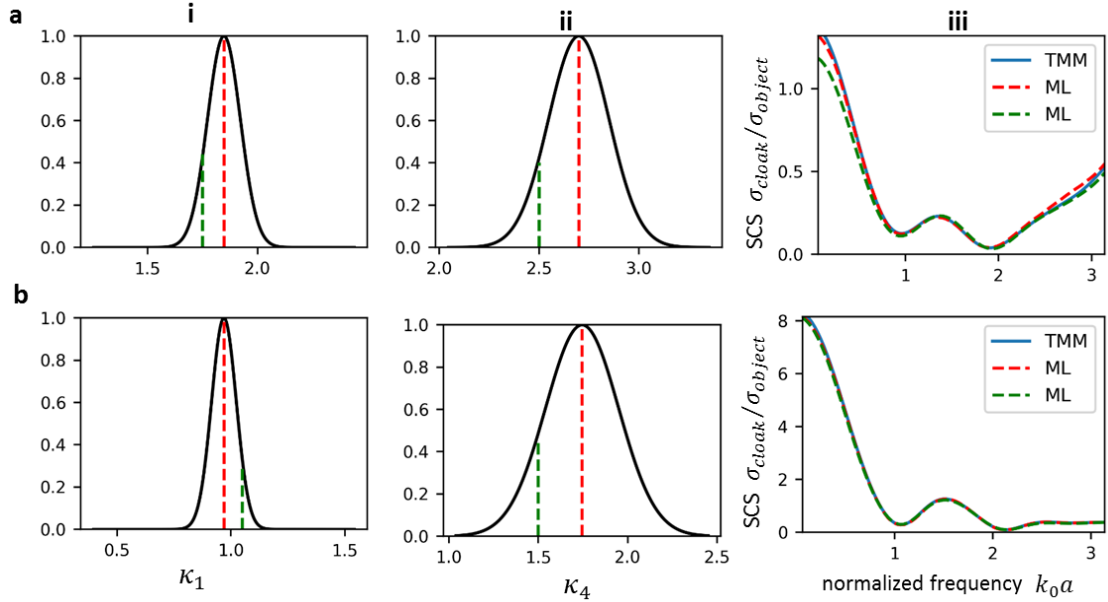

Fig. S6. Examples for probabilistic inverse design network showing maximum deviation in generated distributions of $\kappa_1$ and $\kappa_4$. The parameters are provided in the Table S7.

Table S7. Design parameters for the predicted response in the Fig.S6

| case | $a_1$ | $a_2$ | $a_3$ | $a_4$ | $\kappa_1$ | $\kappa_2$ | $\kappa_3$ | $\kappa_4$ |
|---|---|---|---|---|---|---|---|---|
| Fig.S6a | 1.13 | 1.379 | 1.65 | 1.765 | 1.848 | 0.675 | 0.932 | 2.699 |
| Fig. S6b | 1.024 | 1.528 | 1.659 | 1.727 | 0.969 | 1.107 | 1.865 | 1.745 |

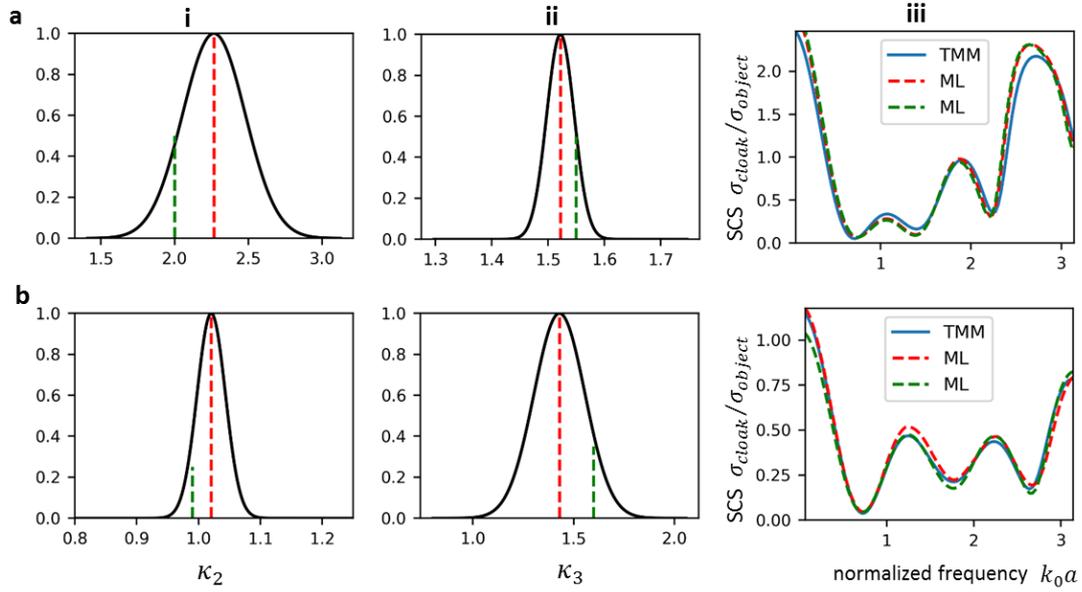

Fig. S7. Examples for probabilistic inverse design network showing maximum deviation in generated distributions of $\kappa_2$ and $\kappa_3$. The parameters are provided in the Table S9.

Table S8. Design parameters for the predicted response in the Fig.S7

| case | $a_1$ | $a_2$ | $a_3$ | $a_4$ | $\kappa_1$ | $\kappa_2$ | $\kappa_3$ | $\kappa_4$ |
|---|---|---|---|---|---|---|---|---|
| Fig. S7a | 1.161 | 1.304 | 1.619 | 1.982 | 0.449 | 2.266 | 1.521 | 0.611 |
| Fig. S7b | 1.071 | 1.296 | 1.396 | 1.754 | 0.484 | 1.021 | 1.43 | 0.689 |

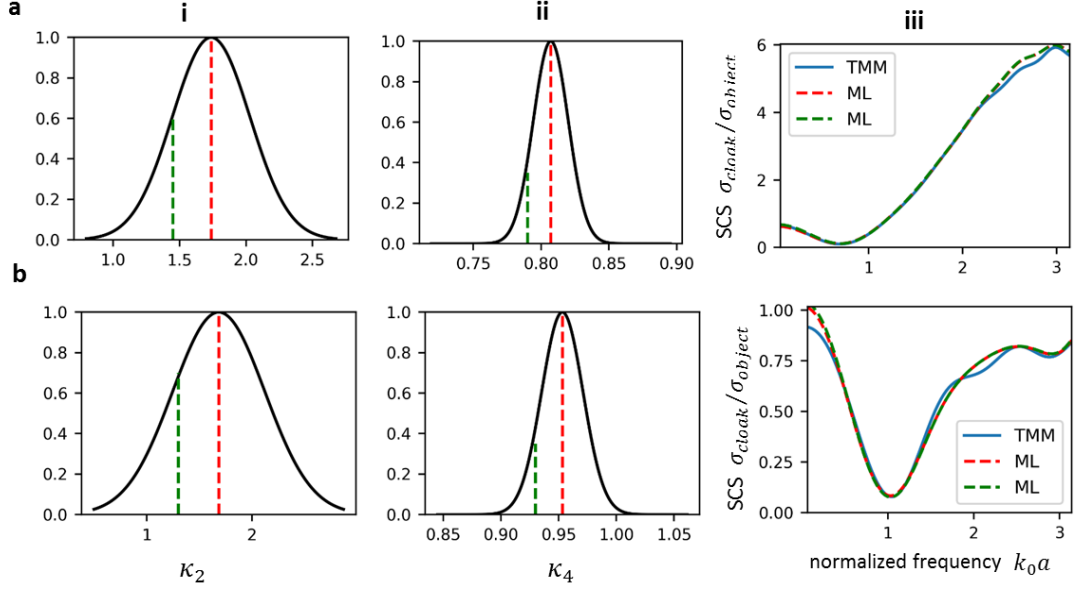

Fig. S8. Examples for probabilistic inverse design network showing maximum deviation in generated distributions of $\kappa_2$ and $\kappa_4$. The parameters are provided in the Table S9.

Table S9. Design parameters for the predicted response in the Fig. S8

| case | $a_1$ | $a_2$ | $a_3$ | $a_4$ | $\kappa_1$ | $\kappa_2$ | $\kappa_3$ | $\kappa_4$ |
|---|---|---|---|---|---|---|---|---|
| Fig. S8a | 1.378 | 1.404 | 1.980 | 2.027 | 2.262 | 1.738 | 0.652 | 0.807 |
| Fig. S8b | 1.189 | 1.196 | 1.378 | 1.428 | 1.918 | 1.683 | 0.391 | 0.953 |

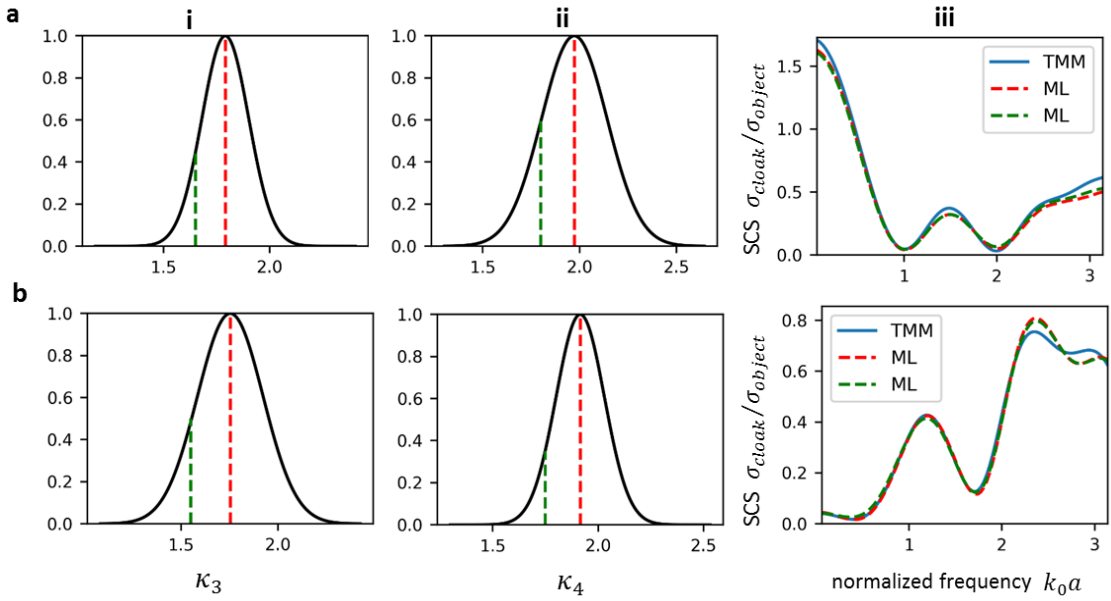

Fig. S9. Examples for probabilistic inverse design network showing maximum deviation in generated distributions of $\kappa_3$ and $\kappa_4$. The parameters are provided in the Table S10.

Table S10. Design parameters for the predicted response in the Fig.S9

| case | $a_1$ | $a_2$ | $a_3$ | $a_4$ | $\kappa_1$ | $\kappa_2$ | $\kappa_3$ | $\kappa_4$ |
|---|---|---|---|---|---|---|---|---|
| Fig. S9a | 1.054 | 1.405 | 1.495 | 1.584 | 0.555 | 0.847 | 1.791 | 1.973 |
| Fig. S9b | 1.159 | 1.475 | 1.512 | 1.577 | 0.522 | 0.852 | 1.754 | 1.915 |